\documentclass[lettersize,journal]{IEEEtran}
\usepackage{amsmath,amsfonts}
\usepackage{amssymb}
\usepackage{algorithmic}
\usepackage{array}
\usepackage{subfig}
\usepackage{textcomp}
\usepackage[short]{optidef}
\usepackage{stfloats}
\usepackage{url}
\usepackage{bm}
\usepackage{verbatim}
\usepackage{comment}
\usepackage{xcolor}
\usepackage{makecell}
\usepackage{booktabs}
\usepackage{adjustbox}
\usepackage{multirow}
\usepackage{cite}
\usepackage{graphicx}
\usepackage[linesnumbered,ruled,vlined]{algorithm2e}
\usepackage{setspace}
\DeclareMathOperator{\Tr}{Tr}
\def\BibTeX{{\rm B\kern-.05em{\sc i\kern-.025em b}\kern-.08em
    T\kern-.1667em\lower.7ex\hbox{E}\kern-.125emX}}
\SetKwInput{KwInput}{Input}                
\SetKwInput{KwOutput}{Output}              
\SetKw{KwInit}{Initialize:}
\usepackage{balance}
\begin{document}
\title{A Meta-Learning Based Precoder Optimization Framework for Rate-Splitting Multiple Access}
\author{Rafael~Cerna~Loli,~\IEEEmembership{Graduate~Student~Member,~IEEE,} and Bruno~Clerckx,~\IEEEmembership{Fellow,~IEEE}
\thanks{R. Cerna Loli is supported by a grant provided by the Defence Science and Technology Laboratory (Dstl) Communications and Networks Research Programme.}
\thanks{R. Cerna Loli and B. Clerckx are with the Department of Electrical and Electronic Engineering, Imperial College London, London SW7 2AZ, U.K. B. Clerckx is also with Silicon Austria Labs (SAL), Graz A-8010, Austria (email: rafael.cerna-loli19@imperial.ac.uk; b.clerckx@imperial.ac.uk).}}

\maketitle

\begin{abstract}
In this letter, we propose the use of a meta-learning based precoder optimization framework to directly optimize the Rate-Splitting Multiple Access (RSMA) precoders with partial Channel State Information at the Transmitter (CSIT). By exploiting the overfitting of the compact neural network to maximize the explicit Average Sum-Rate (ASR) expression, we effectively bypass the need for any other training data while minimizing the total running time. Numerical results reveal that the meta-learning based solution achieves similar ASR performance to conventional precoder optimization in medium-scale scenarios, and significantly outperforms sub-optimal low complexity precoder algorithms in the large-scale regime.
\end{abstract}


\section{Introduction}
\IEEEPARstart{R}{ate-Splitting} Multiple Access (RSMA), has emerged in recent years as a promising multi-antenna multiple access framework for 6G and beyond communications due to its flexible, robust and adaptive interference management capabilities, specially in the presence of partial Channel State Information at the Transmitter (CSIT) \cite{rs_overview}. The benefits of RSMA stem from the fact that it splits the user messages into common parts, decoded by multiple users, and private parts, decoded only by individual users after removing the interference from the common parts using Successive Interference Cancellation (SIC). In this way, RSMA effectively manages interference by partially decoding it, and partially treating it as noise \cite{eurasip}, and, thus, represents the bridge that generalizes other multiple access and transmission strategies, such as Space Division Multiple Access (SDMA), which fully treats interference as noise, Non-Orthogonal Multiple Access (NOMA), which fully decodes interference, Orthogonal Multiple Access (OMA), which avoids interference by transmitting with orthogonal radio resources, and physical-layer multicasting \cite{wcl_2020}.

Due to the multi-antenna nature of RSMA communication systems, precoder optimization plays a fundamental role in managing the interference. Therefore, several works dealing with precoder optimization for RSMA exist in the literature. Regarding RSMA communications with partial CSIT, the authors in \cite{joudeh} proposed the adaptation of the classical Weighted Minimum Mean Square Error (WMMSE) algorithm using the Sampled Average Approximation (SAA) method when the CSIT and the CSIT error distribution are known. However, the application of the SAA-WMMSE optimization is limited by its complexity as the number of antennas and users increase. To overcome this, sub-optimal precoder solutions with low complexity, in which the precoders are designed with fixed direction (e.g. Zero Forcing) and only the power allocation is optimized, are typically used in massive MIMO scenarios \cite{mingbo}. 

In recent years, deep learning (DL) has become an attractive research area in wireless communications. Specifically for DL-based precoder optimization techniques, a \textit{black-box} generic Neural Network (NN) can be trained using labeled data obtained from the WMMSE algorithm in a supervised learning manner \cite{learn_opt}. Another strategy is to employ a \textit{deep-unfolding}-based neural NN with a tailored structure built to replicate the structure of the WMMSE algorithm \cite{learn_opt_3} and reduce the complexity compared to the \textit{black-box} approach. However, these solutions often require large datasets to approach the performance of the WMMSE algorithm. In this letter, inspired by the recent works on \textit{meta-learning} based precoder design for SDMA with perfect CSIT in \cite{lagd_1, lagd_2}, we propose the use of a Meta-Learning Based Precoder Optimization (MLBPO) framework to directly solve the NP-hard precoder optimization problem for RSMA communications with partial CSIT. Unlike the deterministic problems in \cite{lagd_1, lagd_2}, the RSMA precoder optimization problem with partial CSIT is of stochastic nature, and highly non-convex due to the multiple common stream rate constraints. To solve it, we employ a compact NN that is intentionally overfitted to the available CSIT to effectively turn the training phase into a non-convex precoder optimization process. Through numerical results, we demonstrate that MLBPO framework provides similar performance to the SAA-WMMSE optimization, vastly outperforms sub-optimal low complexity precoder solutions in the large-scale massive MIMO regime, and greatly reduces the running time complexity of the precoder optimization process.

\section{System Model}
In this section, we describe the operation of RSMA with Hierarchical Rate-Splitting (HRS) \cite{mingbo}, and with 1-Layer Rate-Splitting (1LRS) \cite{eurasip, joudeh}, which is a special scenario of HRS. 

\subsection{Hierarchical Rate-Splitting Transmission}
We consider a single transmitter equipped with $N_t$ transmit antennas that serves $K$ single-antenna communication users, indexed by the set $\mathcal{K}=\{1,\dots,K\}$\footnote{Due to lack of space, we refer the reader to Fig. 8 in \cite{rs_overview} for a system model figure.}. Additionally, we consider that certain subsets of users possess a degree of similarity (e.g. through spatial correlation) between their channels and, thus, can be grouped accordingly. Assuming that the $K$ users are partitioned into $G$ groups, indexed by the set $\mathcal{G}=\{1,\dots, G\}$, we denote the number of users in the $g$-th group by $K_g$, internally indexed by the set $\mathcal{K}_g=\{1,\dots, K_g\}$. With HRS transmission, the message of user-$k$, $W_k$, is split into a global common part $W_{c,k}$, a group common part $W_{c,g,k}$, and a private part $W_{p,k}$, $\forall k \in \mathcal{K}$. The global common parts of all $K$ users $\{W_{c,1}, \dots, W_{c,K}\}$ are jointly encoded and modulated into a single global common stream $s_c$, the group common parts of the $K_g$ users $\{W_{c,g,1}, \dots, W_{c,g,K_g}\}$ in group-$g$ are jointly encoded, $\forall g \in \mathcal{G}$, and the private parts $\{W_{p,1}, \dots, W_{p,K}\}$ are encoded and modulated independently into $K$ private streams $\{s_1, \dots, s_K\}$. The streams are next linearly precoded using the precoder $\mathbf{P} = [\mathbf{p}_c,\mathbf{p}_{c,1},\dots,\mathbf{p}_{c,G},\mathbf{p}_1,\dots,\mathbf{p}_K] \in \mathbb{C}^{N_t \times (K+G+1)}$, where $\mathbf{p}_c$ is the global common stream precoder, $\mathbf{p}_{c,g}$ is the group common stream precoder for group-$g$, and $\mathbf{p}_k$ is the private stream precoder for user-$k$.  The transmitted signal $\mathbf{x} \in \mathbb{C}^{N_t \times 1}$ is then given by
\begin{equation}
    \mathbf{x} = \mathbf{P}\mathbf{s} = \mathbf{p}_cs_c + \sum_{g=1}^{G}\mathbf{p}_{c,g}s_{c,g} + \sum_{k=1}^{K}\mathbf{p}_ks_k,
    \label{rsma_transmit_signal}
\end{equation}
where $\mathbf{s} = [s_c,s_{c,1},\dots,s_{c,G},s_1,\dots,s_K]^T \in \mathbb{C}^{(K+G+1)\times 1}$. It is assumed that $\mathbb{E}\{\mathbf{s}\mathbf{s}^H\}=\mathbf{I}_{(K+G+1)}$ and, hence, the total transmit power constraint is expressed as $\Tr(\mathbf{P}\mathbf{P}^H)\leq P_t$. The received signal at the output of the antenna of user-$k$, which belongs to group-$g$, is then given by
\begin{equation}
    \begin{split}
        y_k &= \mathbf{h}_k^H\mathbf{p}_cs_c+ \mathbf{h}_k^H\mathbf{p}_{c,g}s_{c,g}+\mathbf{h}_k^H\mathbf{p}_k s_k+\\
        &\;\;\;\underbrace{\sum_{n\neq g}^G\mathbf{h}_k^H\mathbf{p}_{c,n} s_{c,n}}_{\text{inter-group interference}}+\underbrace{\sum_{j\neq k}^K\mathbf{h}_k^H\mathbf{p}_j s_j}_{\text{multi-user interference}}+n_k,
    \end{split}
\end{equation}
where $\mathbf{h}_k \in \mathbb{C}^{N_t \times 1}$ is the downlink channel between the transmitter and user-$k$, and $n_k\;\mathtt{\sim}\; \mathcal{CN}(0,\sigma_{n,k}^2)$ is the Additive White Gaussian Noise (AWGN) at user-$k$.

Decoding at user-$k$ is performed as follows. User-$k$ first performs decoding of the global common stream $s_c$ by treating all group common and private streams as noise. It then subtracts the interference from the global common stream from $y_k$ by employing SIC, and decodes its group common message $s_{c,g}$ by treating the rest of the group common and private streams as noise. Finally, it subtracts the interference of its group common message from the remaining signal by applying SIC again, and decodes its private message $s_k$. The SINRs at user-$k$ of decoding $s_c, s_{c,g}$ and $s_k$, respectively, are given by
\begin{equation}
    \begin{split}
        \gamma_{c,k} &= \frac{|\mathbf{h}_k^H\mathbf{p}_c|^2}{\sum_{n\in\mathcal{G}}^G|\mathbf{h}_k^H\mathbf{p}_{c,n}|^2+\sum_{j\in\mathcal{K}}^K|\mathbf{h}_k^H\mathbf{p}_j|^2+\sigma_{n,k}^2},\\
        \gamma_{c,g,k} &= \frac{|\mathbf{h}_k^H\mathbf{p}_{c,g}|^2}{\sum_{n\neq g}^G|\mathbf{h}_k^H\mathbf{p}_{c,n}|^2+\sum_{j\in\mathcal{K}}^K|\mathbf{h}_k^H\mathbf{p}_j|^2+\sigma_{n,k}^2},\\
        \gamma_{p,k} &= \frac{|\mathbf{h}_k^H\mathbf{p}_k|^2}{\sum_{n\neq g}^G|\mathbf{h}_k^H\mathbf{p}_{c,n}|^2+\sum_{j\neq k}^K|\mathbf{h}_k^H\mathbf{p}_j|^2+\sigma_{n,k}^2}.
    \end{split}
\end{equation}

At user-$k$, and assuming Gaussian signalling, the achievable rate of the global common stream is $R_{c,k} = \log_2(1+\gamma_{c,k})$, the achievable rate of the group common stream of group-$g$ is given by $R_{c,g,k}=\log_2(1+\gamma_{c,g,k})$, and the achievable rate of its private stream is $R_{k} = \log_2(1+\gamma_{p ,k})$. As the global common stream and group common streams must be decoded by more than one user, they must be transmitted respectively at rates not exceeding $R_c = \min\{R_{c,1},\dots,R_{c,K}\}$ and $R_{c,g}=\min\{R_{c,g,1},\dots,R_{c,g,K_g}\}, \forall g \in \mathcal{G}$. The HRS system Sum-Rate (SR) expression is then given by 
\begin{equation}
    \text{SR}_{\text{HRS}}(\mathbf{P}) = R_c + \sum_{g=1}^G R_{c,g} + \sum_{k=1}^K R_k.
    \label{hrs_sr}
\end{equation}

\subsection{1-Layer Rate-Splitting Transmission}
In 1LRS transmission, only the global common stream and $K$ private streams are scheduled. Thus, 1LRS is a special case of HRS in which the group common stream precoders are deactivated (i.e. setting $\mathbf{p}_{c,g}=\bm{0}^{N_t \times 1}$, $\forall g \in \mathcal{G}$), and the 1LRS system SR expression is a reduced version of (\ref{hrs_sr}) given by
\begin{equation}
    \text{SR}_{\text{1LRS}}(\mathbf{P}) = R_c + \sum_{k=1}^K R_k.
    \label{1lrs_sr}
\end{equation}

\subsection{Channel State Information Model}
Considering partial CSIT is a realistic assumption as several factors (e.g. quantized feedback, feedback delays) can degrade the CSIT quality in a practical system. Therefore, the CSI model is given by \cite{lina_dpc}
\begin{equation}
    \mathbf{H}=\mathbf{\hat{H}}+\mathbf{\Tilde{H}},
\end{equation}
where $\mathbf{H}=[\mathbf{h}_1,\dots,\mathbf{h}_K]$ is the real CSI with i.i.d elements drawn from the distribution $\mathcal{CN}(0,\sigma_k^2), \forall k\in \mathcal{K}$, and $\sigma_k^2$ being the channel amplitude power. Also, $\hat{\mathbf{H}}=[\hat{\mathbf{h}}_1,\dots,\hat{\mathbf{h}}_K]$ is the CSIT with the elements of $\hat{\mathbf{h}}_k$ following a distribution $\mathcal{CN}(0,\sigma_k^2-\sigma_{e,k}^2), \forall k\in \mathcal{K}$. Finally, $\Tilde{\mathbf{H}}=[\Tilde{\mathbf{h}}_1,\dots,\Tilde{\mathbf{h}}_K]$ represents the CSI estimation error, with the elements of $\Tilde{\mathbf{h}}_k$ following a distribution $\mathcal{CN}(0,\sigma_{e,k}^2), \forall k\in \mathcal{K}$. The parameter $\sigma_{e,k}^2$ is defined as the CSIT error power for user-$k$. The perfect CSIT scenario can then be represented by choosing $\sigma_{e,k}^2 = 0$.

\section{Meta-Learning Based Precoder Optimization}
In this section, we first describe the HRS precoder optimization problem formulation with partial CSIT, and then we introduce the proposed MLBPO framework to solve it.
\subsection{Precoder optimization problem formulation}
Computing the optimum precoders that maximize $(\ref{hrs_sr})$ and $(\ref{1lrs_sr})$ is not possible due to the CSIT error uncertainty. To overcome this, it was proposed in \cite{joudeh} that a more robust approach with partial CSIT is to optimize the precoders to maximize the Ergodic Rates (ERs) of each stream. This, in turn, can be achieved by maximizing the Average Rates (ARs), which represent the short-term expected rates over the conditional error distribution $f_{\text{H}|\hat{\text{H}}}(\mathbf{H}|\hat{\mathbf{H}})$, of each stream over a sufficiently large set of random CSIT realizations $\mathbf{\hat{H}}$. Thus, the global common, group common and private ARs of user-$k$ are given respectively by $\bar{R}_{c,k}\triangleq\mathbb{E}_{\text{H}|\hat{\text{H}}}\{R_{c,k}|\hat{\mathbf{H}}\}$, $\bar{R}_{c,g,k}\triangleq\mathbb{E}_{\text{H}|\hat{\text{H}}}\{R_{c,g,k}|\hat{\mathbf{H}}\}$ and  $\bar{R}_k\triangleq\mathbb{E}_{\text{H}|\hat{\text{H}}}\{R_{k}|\hat{\mathbf{H}}\}$. The HRS system Average SR (ASR) is then given by
\begin{equation}
    \text{ASR}_{\text{HRS}}(\mathbf{P}) = \bar{R}_c + \sum_{g=1}^G \bar{R}_{c,g} + \sum_{k=1}^K \bar{R}_k.
\end{equation}

To turn the stochastic ASR expression into a deterministic one for a given CSIT $\mathbf{\hat{H}}$ and $f_{\text{H}|\hat{\text{H}}}(\mathbf{H}|\hat{\mathbf{H}})$, we can employ the Sample Average Approximation method to estimate the ARs of each stream. Therefore, we first generate a set of $M$ i.i.d CSIT error realizations, indexed by the set $\mathcal{M}\triangleq\{1,\dots,M\}$, for a given CSIT error variance $\sigma_{e}^2$, given by $\Tilde{\mathbb{H}}^{(\textbf{M})}\triangleq\{\Tilde{\mathbf{H}}^{(m)}|\;m \in \mathcal{M}\}$. The ensemble of $M$ real CSI realizations associated to the CSIT error set $\Tilde{\mathbb{H}}^{(\textbf{M})}$ is given by  
\begin{equation}
    \mathbb{H}^{(\textbf{M})}\triangleq\{\mathbf{H}^{(m)}=\hat{\mathbf{H}}+\Tilde{\mathbf{H}}^{(m)}|\;\hat{\mathbf{H}},\;m \in \mathcal{M}\}.
\end{equation}
From the strong Law of Large Numbers, the ARs of each stream can be estimated through their Sample Average Functions (SAFs) as $M \rightarrow \infty$. The SAFs are characterized by $\bar{R}_{c,k}^{(\textbf{M})}\triangleq\frac{1}{M}\sum_{m=1}^MR_{c,k}^{(m)}$, $\bar{R}_{c,g,k}^{(\textbf{M})}\triangleq\frac{1}{M}\sum_{m=1}^MR_{c,g,k}^{(m)}$, and $\bar{R}_k^{(\textbf{M})}\triangleq\frac{1}{M}\sum_{m=1}^MR_k^{(m)}$, where $R_{c,k}^{(m)}$, $R_{c,g,k}^{(m)}$, and $R_k^{(m)}$ are the achievable rates associated with the global common, group common, and private streams at user-$k$ for the $m$-th CSI realization $\mathbf{H}^{(m)}$ in the ensemble $\mathbb{H}^{(M)}$. The SAA of the ASR maximization is then expressed as
\begin{maxi} |s|[2]
{\mathbf{P}}{\text{ASR}_{\text{HRS}}^{(\mathbf{M})}(\mathbf{P}) = \bar{R}_{c}^{(\textbf{M})}+\sum_{g=1}^G\bar{R}_{c,g}^{(\textbf{M})}+\sum_{k=1}^K\bar{R}_{k}^{(\textbf{M})}}{\label{saa_opt_prob}}{} 
\addConstraint{\bar{R}_{c}^{(\textbf{M})}}{\leq\bar{R}_{c,k}^{(\textbf{M})},\quad}{\forall k\in\mathcal{K}}
\addConstraint{\bar{R}_{c,g}^{(\textbf{M})}}{\leq\bar{R}_{c,g,k_g}^{(\textbf{M})},\quad}{\forall k_g\in\mathcal{K}_g\;,\; g\in\mathcal{G}}
\addConstraint{\Tr(\mathbf{P}\mathbf{P}^H)}{\leq P_t,}
\end{maxi}
where $\mathbf{P}$ is fixed for all CSI realizations in the ensemble $\mathbb{H}^{(M)}$.

\subsection{Proposed Solution}
Among the conventional non-learning approaches to solve the NP-hard non-convex optimization problem in ($\ref{saa_opt_prob}$), the SAA Weighted Minimum Mean Square Error (WMMSE) optimization algorithm, that transforms the problem in ($\ref{saa_opt_prob}$) into a Quadratically Constrained Quadratic Program (QCQP) which can be solved using convex optimization tools, is by far the most well-known \cite{joudeh}. However, it suffers from extremely high time complexity in the order of $\mathcal{O}(L(N_tK)^{3.5})$ \cite{complex_1}, where $L$ denotes the number of iterations the algorithm runs for. Thus, employing it in scenarios with $N_t\gg 1$ and $K\gg1$ requires an exponentially large and impractical running time.

Inspired by the recent works on meta-learning based non-convex optimization \cite{mlam}, and learning-aided gradient descent for MU-MISO \cite{lagd_1} and MU-MIMO \cite{lagd_2} beamforming with perfect CSIT, we propose a meta-learning based precoder optimization framework to directly solve ($\ref{saa_opt_prob}$), which employs the current CSIT $\mathbf{\hat{H}}$ as the sole training data. To achieve this, a single compact NN, denoted by $\text{G}_{\bm{\theta}}(.)$, is intentionally overfitted to $\mathbf{\hat{H}}$ during training in order to have its tunable parameters, denoted by $\bm{\theta}$, learn a meta-learning, adaptive precoder update rule specific to $\mathbf{\hat{H}}$ that minimizes the following loss function in an unsupervised learning manner
\begin{equation}
    \mathcal{L}(\mathbf{P})=-\text{ASR}_{\text{HRS}}^{(\mathbf{M})}(\mathbf{P}).
\end{equation}
The general structure of the MLBPO framework is presented in \textbf{Algorithm 1}, and a detailed description of it is given next.
\begin{algorithm}[!t]
\DontPrintSemicolon
  \KwInput{$N_t, K, \mathbf{P}_0, \mathbb{H}^{(\textbf{M})}, L, \beta.$}
  \KwInit{$\bm{\theta}_0$}
  
  \For{$i \leftarrow 0,1,\dots,L-1$}
  {$\mathbf{P}_{i+1} = \mathbf{P}_0 + \text{G}_{\bm{\theta}_i}(\nabla_{\mathbf{P}_0}\mathcal{L}(\mathbf{P}_0))$ \;
  $\mathbf{P}_{i+1} = \Omega(\mathbf{P}_{i+1})$\;
  $\bm{\theta}_{i+1} = \bm{\theta}_{i} +\beta \cdot \text{Adam}(\nabla_{\bm{\theta}_{i}}\mathcal{L}(\mathbf{P}_{i+1}))$ \;
  }
  \KwOutput{$\mathbf{P}_{L}$}
\caption{MLBPO for ASR maximization}
\label{radcom_admm_algorithm_imperfect}
\end{algorithm}

At the $i$-th iteration, the network $\text{G}_{\bm{\theta}_i}(.)$ takes as input the gradient $\nabla_{\mathbf{P}_0}\mathcal{L}(\mathbf{P}_0)$, where $\mathcal{L}(\mathbf{P}_0)$ is the loss achieved using the initial precoder $\mathbf{P}_0$, a sub-optimal estimation designed as a function of $\mathbf{\hat{H}}$. The network $\text{G}_{\bm{\theta}_i}(.)$ then outputs the incremental precoder update term to update $\mathbf{P}_0$ as follows
\begin{equation}
    \mathbf{P}_{i+1} = \mathbf{P}_0 + \text{G}_{\bm{\theta}_i}(\nabla_{\mathbf{P}_0}\mathcal{L}(\mathbf{P}_0)).
\end{equation}
As previously mentioned, this incremental precoder update strategy based on the fixed initial point $\mathbf{P}_0$ is performed to exploit the overfitting of the NN $\text{G}_{\bm{\theta}}(.)$ and, hence, effectively turning the unsupervised training phase into a non-linear, non-convex optimization process that can directly solve ($\ref{saa_opt_prob}$). $\mathbf{P}_{i+1}$ is then projected to comply with the total transmit power constraint $\Tr(\mathbf{P}\mathbf{P}^H)\leq P_t$ according to
\begin{equation}
\Omega(\mathbf{P}) = 
\begin{cases}
        \mathbf{P}, & \text{if } \Tr(\mathbf{P}\mathbf{P}^H)\leq P_t\\
        \sqrt{\frac{P_t}{\Tr({\mathbf{P}\mathbf{P}^H})}}\mathbf{P} & \text{otherwise.}
    \end{cases}
\end{equation}
Finally, the update term for $\bm{\theta}$ is obtained through employing the Adam \cite{adam} optimizer with respect to the loss achieved by the updated precoder $\mathbf{P}_{i+1}$ as follows \cite{mlam}
\begin{equation}
    \bm{\theta}_{i+1} = \bm{\theta}_{i} + \beta \cdot \text{Adam}(\nabla_{\bm{\theta}_{i}}\mathcal{L}(\mathbf{P}_{i+1})),
\end{equation}
where $\beta$ is the learning rate parameter of the Adam optimizer. 

We then summarize the main advances of the proposed MLBPO
framework:
\begin{itemize}
    \item \textit{First, } the proposed solution is able to exploit the non-linearity of the compact NN to directly optimize ($\ref{saa_opt_prob}$), compared to having to rely on sub-optimal convex relaxation techniques, or alternatively settle for sub-optimal low complexity solutions that only optimize the power allocated to each precoder in $\mathbf{P}$.
    \item \textit{Second, } the proposed solution offers a substantially lower complexity in the order of $\mathcal{O}(LN_t(K+G+1))$ as it only depends on the number of elements in $\mathbf{P}$. Thus, it can be applied to large-scale scenarios with $N_t\gg 1$ and $K\gg1$.    
    \item \textit{Third, } by employing the meta-learning strategy for each $\mathbf{\hat{H}}$, the proposed solution effectively avoids the need for long training times and large training datasets as the compact NN is retrained and overffited for each $\mathbf{\hat{H}}$.
\end{itemize}

\section{Numerical Results}
In this section, we compare the 1LRS and HRS Ergodic SR (ESR), average running time, and average precoder power allocation of the MLBPO framework with the SAA-WMMSE optimization and the sub-optimal low complexity solution in \cite{mingbo}. All results are obtained by averaging the results over 100 random CSIT realizations. The channel ensemble $\mathbb{H}^{(\textbf{M})}$ is generated considering $M=1000$. Additionally, we consider that the noise power is 
$\sigma_{n,k}^2=1,\forall k \in \mathcal{K}$. The MLBPO framework is implemented in Python 3.9 with PyTorch 1.12.1, using an NVIDIA RTX 6000 GPU. All other algorithms are implemented in MATLAB R2022b with the CVX toolbox \cite{cvx_web}, using an Intel Xeon Platinum 8358 2.60GHz CPU. 

\subsection{1-Layer Rate-Splitting Transmission}
We first compare the 1LRS ESR performance of the MLBPO framework, and the classical SAA-WMMSE algorithm as the baseline. We consider $N_t=16$ and $K=16$, a medium-scale scenario. Also, we assume that the CSIT error power scales with the SNR as $\sigma_{e,k}^2=P_t^{-\alpha},\forall k \in \mathcal{K}$, where $\alpha=0.6$ is the CSIT scaling factor.

We set the number of iterations for both the MLBPO framework and SAA-WMMSE optimization to $L=500$. Additionally, to minimize the total running time of the SAA-WMMSE optimization, we stop it early when convergence of the ASR is reached. We consider that this occurs when the difference between the achieved ASR in consecutive iterations is less than or equal to $10^{-6}$ bps/Hz. We also employ the same Singular Value Decomposition (SVD) and Maximum Ratio Transmission (MRT) \cite{joudeh} method to initialize $\mathbf{P}_0$, allocating 90\% of $P_t$ to the initial common stream precoder, and the remaining 10\% equally distributed among the $K$ private stream precoders. Finally, the NN $\text{G}_{\bm{\theta}}(.)$ of the MLBPO framework contains two hidden layers with 50 neurons each and the learning rate of the Adam Optimizer is $\beta=10^{-3}$.

\begin{figure}[t!]
\begin{minipage}{.5\linewidth}
\centering
\subfloat[]{\label{main:asr}\includegraphics[scale=.3]{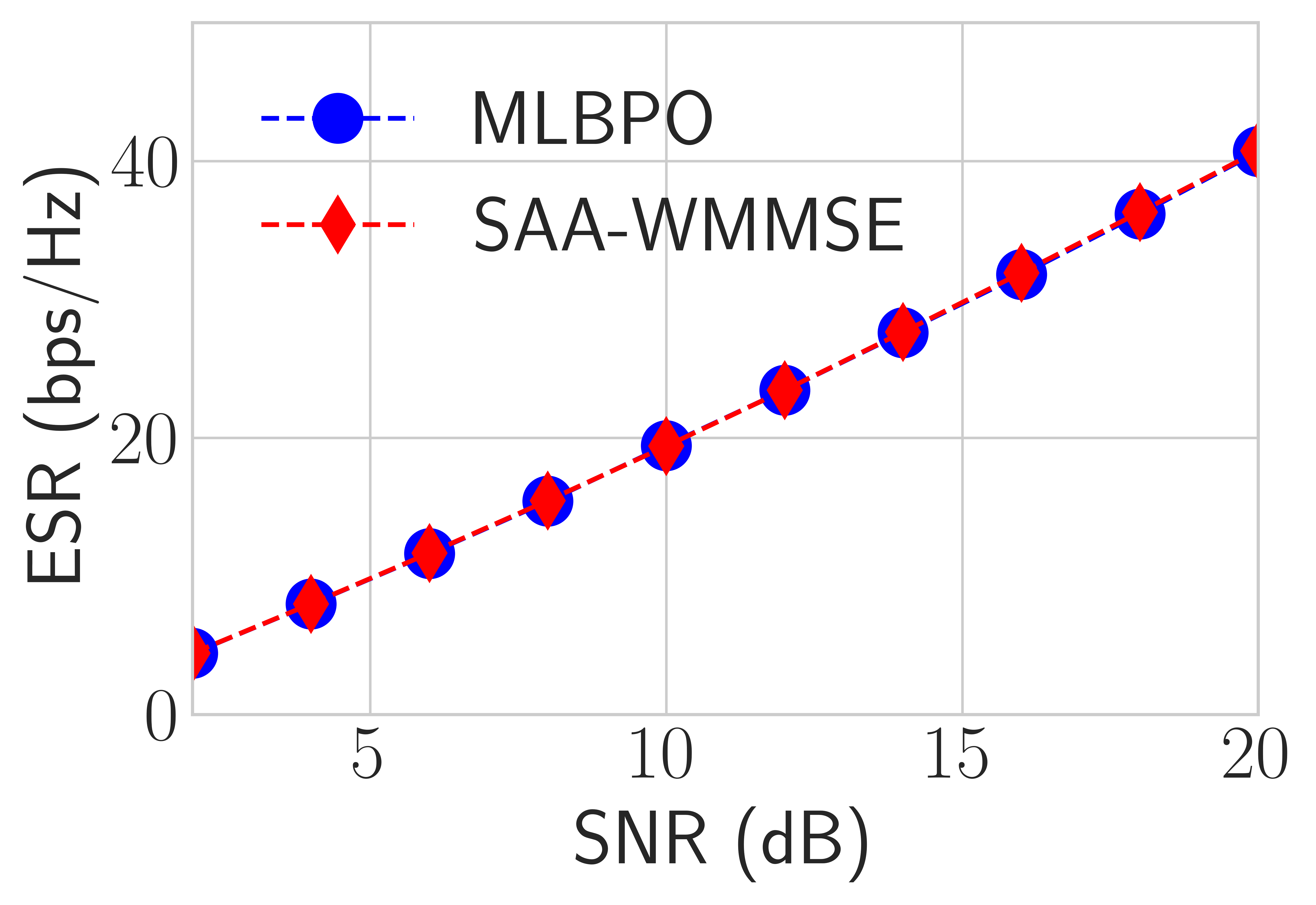}}
\end{minipage}%
\begin{minipage}{.5\linewidth}
\centering
\subfloat[]{\label{main:run_time}\includegraphics[scale=.3]{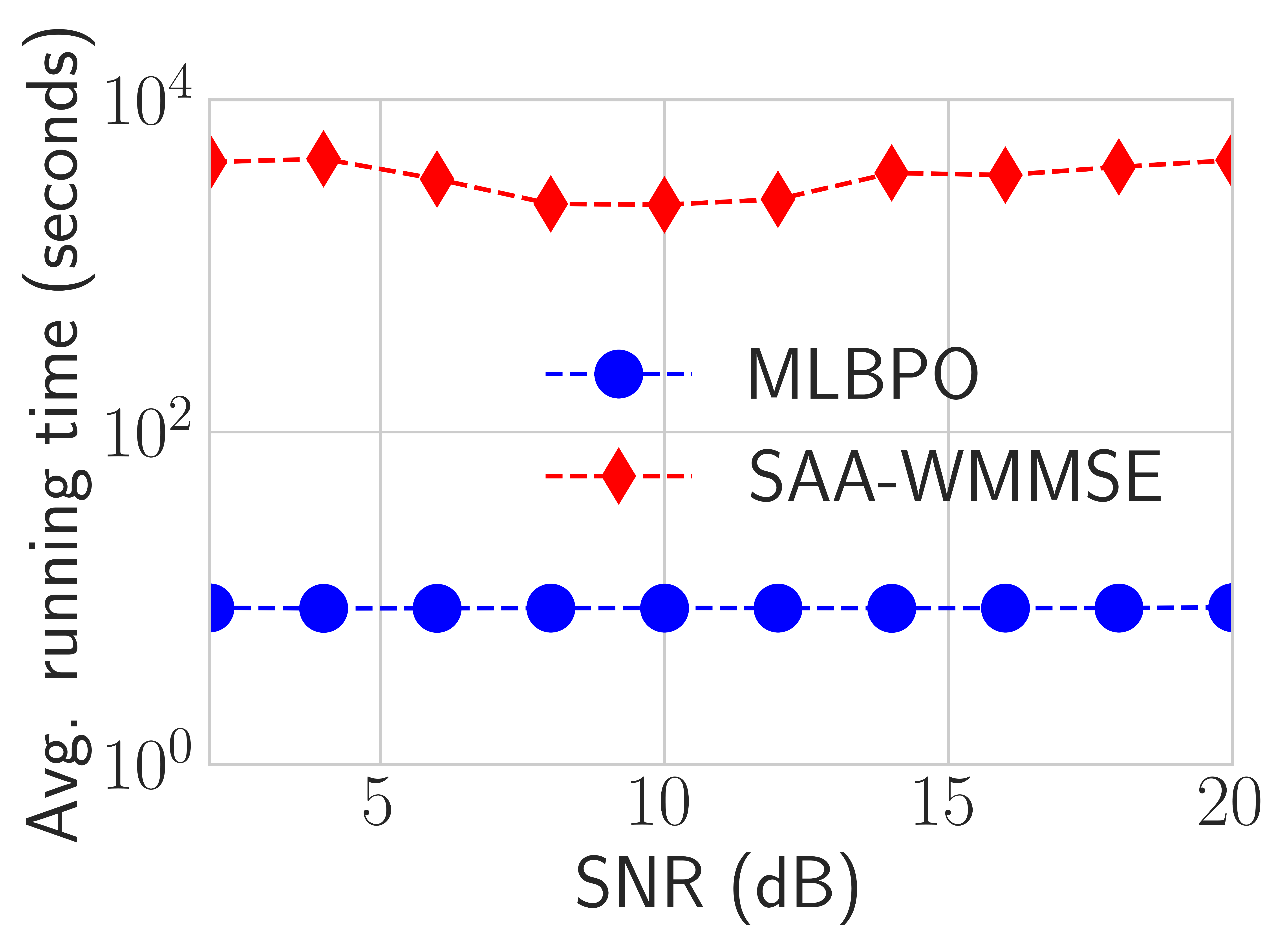}}
\end{minipage}\par\medskip
\caption{1LRS transmission with $N_t=16, K=16$: (a) ESR vs. SNR (b) Avg. Running Time vs. SNR.}
\label{fig:1lrs_perf}
\end{figure}

Results are shown in terms of the ESR vs. SNR and average running time vs. SNR in Fig. \ref{main:asr} and Fig. \ref{main:run_time}, respectively. It can be immediately observed from Fig. \ref{main:asr} that the ESR of the MLBPO framework and the SAA-WMMSE optimization are virtually identical. Nevertheless, the main advantage of the MLBPO framework is revealed from Fig. \ref{main:run_time}, in which it is clearly observed that the average running time of the MLBPO framework is two orders of magnitude lower than the average running time of the SAA-WMMSE optimization in all the SNR range. Specifically, the average running time across all SNR values for the MLBPO framework is 8.72 seconds, a significantly lower time compared to the 3463.92 seconds of the SAA-WMMSE optimization. Thus, this demonstrates that the MLBPO framework stands as a much more practical solution to solving the 1LRS ASR maximization problem than the classical SAA-WMMSE optimization.
\subsection{Hierarchical Rate-Splitting Transmission}
To assess the performance of the MLBPO framework when solving the HRS ASR maximization problem, we compare it with the sub-optimal low complexity precoder solution for HRS in the massive MIMO regime ($N_t \gg K > 1$) that was presented in \cite{mingbo}. We consider $N_t=100$ and $K=12$ equally grouped in $G=4$ groups, located in azimuth directions $[\frac{-\pi}{2}, \frac{-\pi}{6},\frac{\pi}{6},\frac{\pi}{2}]$. It is assumed that the users in each group share the same spatial correlation matrix $\mathbf{R}_g$ obtained by considering a geometrical one-ring scattering model. Thus, the channel of user-$k$ in group-$g$ is expressed as $\mathbf{h}_k=\mathbf{R}_g^{\frac{1}{2}}\mathbf{g}_k$, where $\mathbf{g}_k$ possesses i.i.d entries drawn from the distribution $\mathcal{CN}(0,1)$. The CSIT model in this scenario is given by
\begin{equation}
    \mathbf{\hat{h}}_k = \mathbf{R}_g^{\frac{1}{2}}\Big(\sqrt{1-\tau_k^2}\mathbf{g}_k+\tau_k\mathbf{z}_k\Big),
\end{equation}
where $\mathbf{z}_k$ is the CSIT error with i.i.d entries drawn from the distribution $\mathcal{CN}(0,1)$, and $\tau_k \in [0, 1]$ denotes the instantaneous CSIT quality for user-$k$.

We then consider $\tau_k^2=0.4$ to simulate two different scenarios: a first one in which the user groups are spatially disjoint with angular spread $\Delta = \frac{\pi}{8}$, and a second one in which the user groups are spatially overlapping with angular spread $\Delta = \frac{\pi}{3}$. Finally, the NN $\text{G}_{\bm{\theta}}(.)$ of the MLBPO framework contains three hidden layers with 300 neurons each and the learning rate of the Adam optimizer is $\beta=10^{-4}$. 
\begin{figure}[t!]
\begin{minipage}{.5\linewidth}
\centering
\subfloat[]{\label{main:asr_disjoint}\includegraphics[scale=.3]{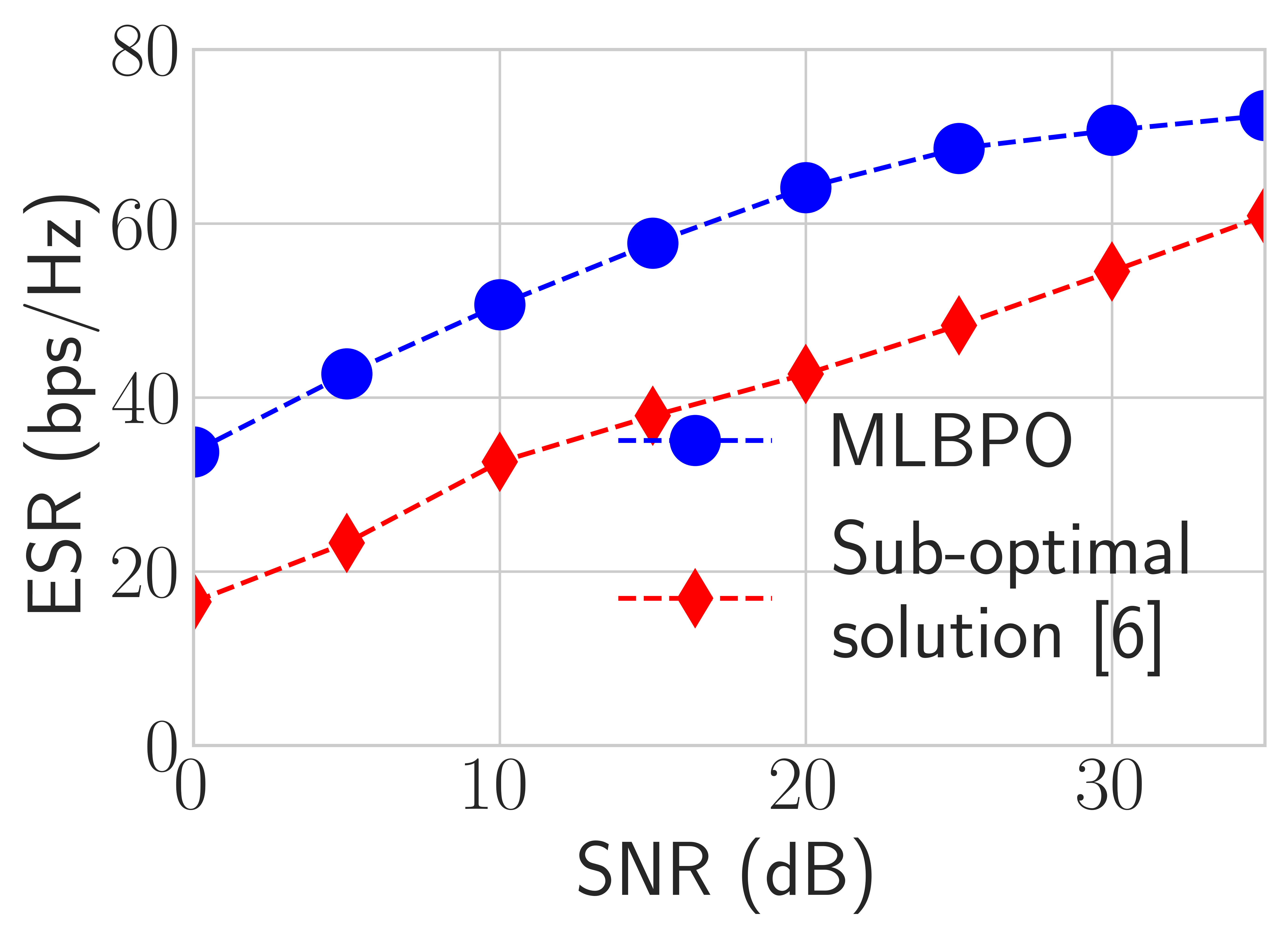}}
\end{minipage}%
\begin{minipage}{.5\linewidth}
\centering
\subfloat[]{\label{main:asr_overlap}\includegraphics[scale=.3]{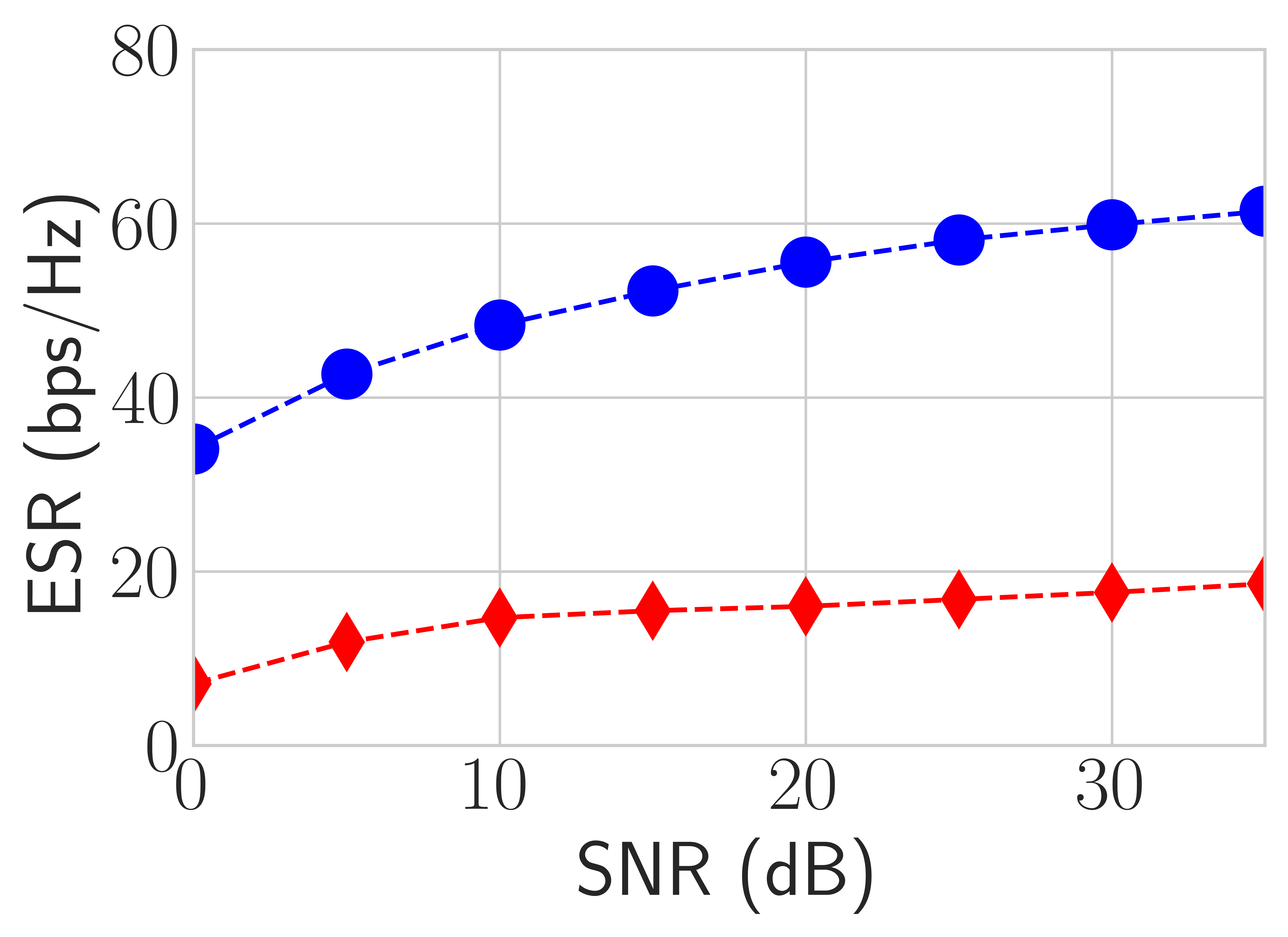}}
\end{minipage}\par\medskip
\caption{HRS transmission with $N_t=100, G=4, K=12$: ESR vs. SNR (a) $\Delta=\frac{\pi}{8}$ (b) $\Delta=\frac{\pi}{3}$.}
\label{fig:hrs_asr}
\end{figure}

\begin{figure}[t!]
\begin{minipage}{.5\linewidth}
\centering
\subfloat[]{\label{main:pow_disjoint}\includegraphics[scale=.3]{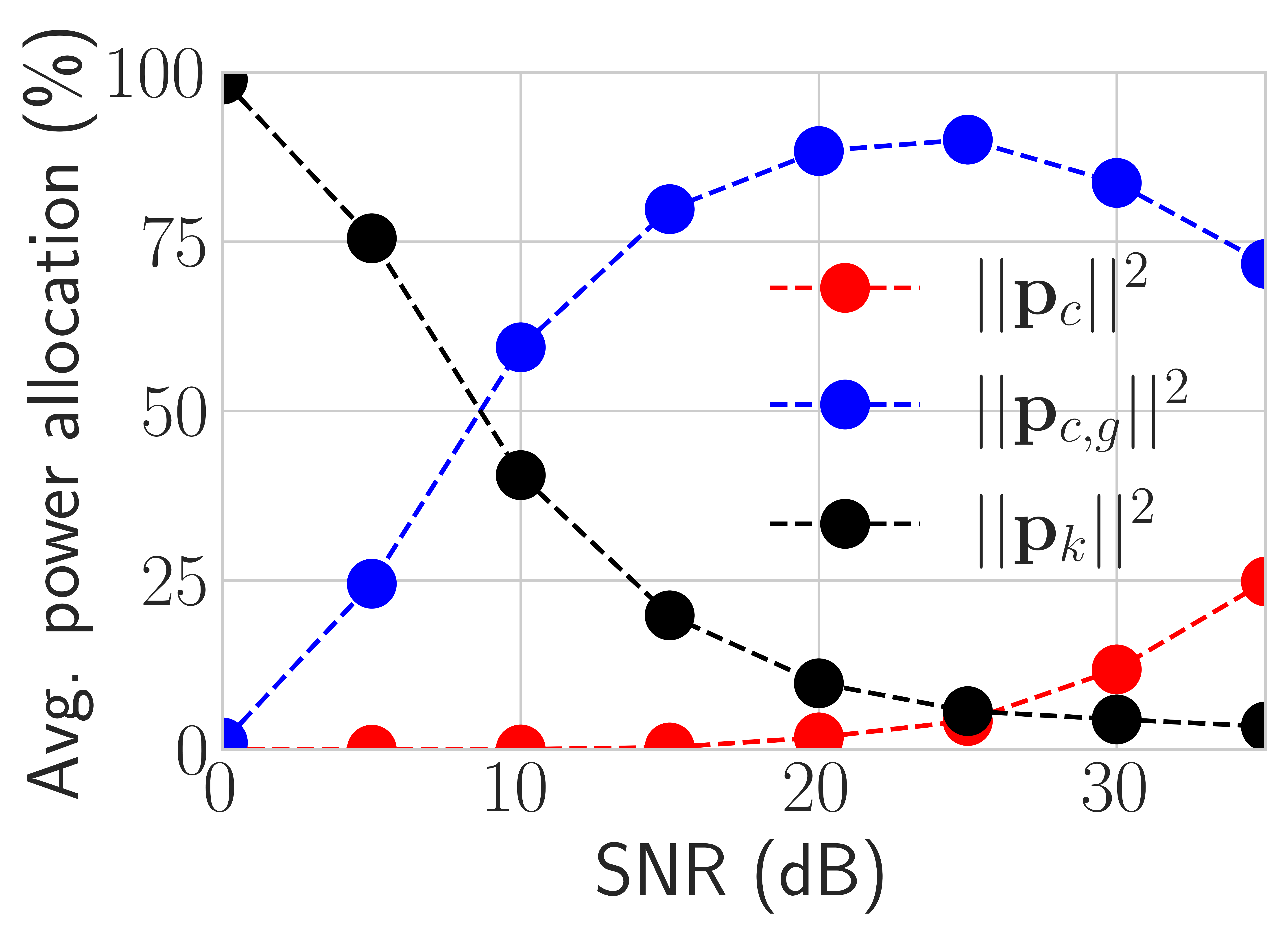}}
\end{minipage}%
\begin{minipage}{.5\linewidth}
\centering          
\subfloat[]{\label{main:pow_overlap}\includegraphics[scale=.3]{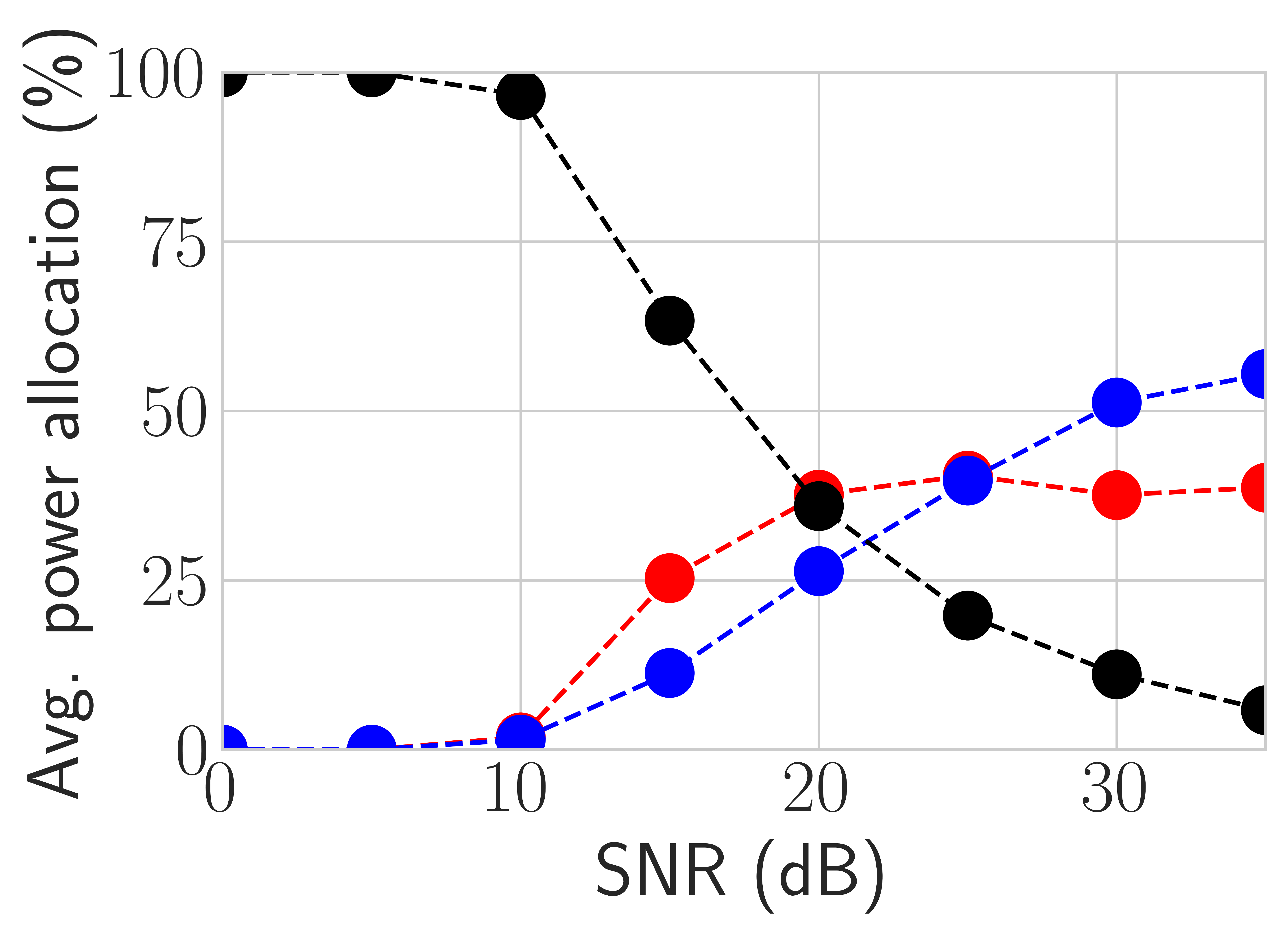}}
\end{minipage}\par\medskip
\caption{$N_t=100, G=4, K=12$: MLBPO average precoder power allocation (a) $\Delta=\frac{\pi}{8}$ (b) $\Delta=\frac{\pi}{3}$.}
\label{fig:hrs_pow}
\end{figure}
Results for the two scenarios are first shown in terms of the ESR vs. SNR in Fig. \ref{fig:hrs_asr} where it can be observed that the MLBPO framework vastly outperforms the sub-optimal precoder solution, especially for $\Delta = \frac{\pi}{3}$. To explain this, it is important to indicate that \cite{mingbo} employs an eigendecomposition-based technique of the correlation matrices $\mathbf{R}_g$ to decompose the total precoder matrix $\mathbf{P}$ into an outer and and inner part with fixed structure, and only the power ratios between global common, group common, and private stream precoders are optimized. Additionally, to reduce the total dimensionality of the problem, the outer and inner precoders are steered only in the directions of the eigenvectors corresponding to a fixed number of the dominant eigenvalues of the channel matrices of each group. In this way, it totally disregards the non-negligible interference in the eigendirections of the vanishing eigenvalues, which ultimately degrades the ESR. In contrast, the MLBPO framework directly optimizes $\mathbf{P}$ without such prior assumptions and converges to a more optimal solution.

 To further illustrate the superiority of the MLBPO framework, we present results in terms of the average precoder power allocation in Fig. \ref{fig:hrs_pow}. According to \cite{mingbo}, for the disjoint group scenario with $\Delta = \frac{\pi}{8}$, the global common stream is deactivated in the low-complexity solution and increasing power to the group common streams is allocated as the SNR increases. Thus, per-group 1LRS transmission is established. The MLBPO framework also follows this trend for SNR levels less than 20 dB, as shown in Fig. \ref{main:pow_disjoint}. However, for higher SNR levels, the MLBPO framework increases the power allocated to the global common stream precoder and decreases the power of the group common stream precoders in order to partially decode inter-group interference and avoid operating in an inter-group interference limited state. Thus, the ESR can continue increasing. In the overlapping group scenario with $\Delta = \frac{\pi}{3}$, \cite{mingbo} proposes that the group common precoders are instead deactivated and only the global common stream should be allocated increased power as the SNR increases. In contrast, it is observed from Fig. \ref{main:pow_overlap} that the MLBPO framework allocates increasing power to the group common precoders. Upon further inspection of the individual group common precoder power allocation, it is revealed that the power is allocated to the group common stream precoders of the groups in directions $[\frac{-\pi}{2}, \frac{\pi}{2}]$, which are the only spatially disjoint groups. Thus, these results reveal that, since the MLBPO framework does not optimize precoders with reduced dimensionality and fixed structure as \cite{mingbo}, it is capable of jointly tuning the individual gains and phases of all elements in the precoder matrix $\mathbf{P}$ more effectively to converge to a more optimal solution that can exploit all three stream categories. Finally, we also indicate that the average running time of the MLBPO framework is 67.53 seconds, and the average running time of the sub-optimal low complexity solution is 0.02 seconds.
 
\section{Conclusion}
We propose a MLBPO framework for RSMA, which fully exploits the overfitting effect of a compact NN to turn the unsupervised training phase into an effective non-linear non-convex optimization process for ASR maximization with partial CSIT. Due to the simplicity of the operation of the compact NN, the MLBPO framework achieves a significant reduction in time complexity compared to the classical SAA-WMMSE optimization algorithm, while achieving very similar ASR performance. In large-scale scenarios, the MLBPO framework demonstrates a substantial ASR gain over other sub-optimal low complexity precoder solutions by jointly optimizing all elements in the precoder matrix.

\end{document}